\documentclass[useAMS,usenatbib]{mnras}
\usepackage{times,graphicx,amsmath,amsfonts,amssymb,aas_macros,epstopdf}
\usepackage{epsfig}
\usepackage[usenames,dvipsnames]{xcolor}
\usepackage{url}
\usepackage{subfigure}
\usepackage[normalem]{ulem} 
\usepackage{tabularx}

\usepackage{array}
\newcolumntype{L}[1]{>{\raggedright\let\newline\\\arraybackslash\hspace{0pt}}m{#1}}
\newcolumntype{C}[1]{>{\centering\let\newline\\\arraybackslash\hspace{0pt}}m{#1}}
\newcolumntype{R}[1]{>{\raggedleft\let\newline\\\arraybackslash\hspace{0pt}}m{#1}}

\def\etal{{\frenchspacing\it et al.}}

\def\ltsim{\lower.5ex\hbox{$\; \buildrel < \over \sim \;$}}
\def\gtsim{\lower.5ex\hbox{$\; \buildrel > \over \sim \;$}}
\def\ltsim{\lower.5ex\hbox{$\; \buildrel < \over \sim \;$}}
\def\gtsim{\lower.5ex\hbox{$\; \buildrel > \over \sim \;$}}
\def\be{\begin{equation}}
\def\ee{\end{equation}}
\def\ba{\begin{eqnarray}}
\def\ea{\end{eqnarray}} 
\def\Msun{h^{-1}{\rm M}_{\odot}}
\def\hmpc{h^{-1}\,{\rm Mpc}}
\def\hgpc{h^{-1}\,{\rm Gpc}}
\def\kms{\, {\rm km }\, {\rm s}^{-1}}
\def\mnras{MNRAS}

\def\dd{\textrm{d}}

\def\fnl{f_{_{\rm NL}}}

\def\ln{{\rm ln}\,}

\def\frac#1#2{{\textstyle{#1\over #2}}}

\def\bx{\mathbf{x}}

\title[Dips and Peaks]{Abundance of peaks and dips in three-dimensional mass and  halo density fields: a test for cosmology}
\author[Nusser \etal]{Adi Nusser$^{1}$\thanks{E-mail: adi@physics.technion.ac.il}, Matteo Biagetti$^{2}$, 
 Vincent Desjacques$^{1}$\\ 
$^{1}$Department of Physics and the Asher Space Research Institute, Israel Institute of Technology Technion, Haifa 32000, Israel\\
$^{2}$Institute of Physics (IoP), Faculty of Science, University of Amsterdam, Science Park 904 | 1098 XH Amsterdam,  the Netherlands
}

\begin{document}
\date{Accepted XXXXXX. Received XXXXXX; in original form XXXXXX}
\pagerange{\pageref{firstpage}--\pageref{lastpage}} \pubyear{2018}
\maketitle
\label{firstpage}

\begin{abstract}

  Using cosmological  N-body simulations, we study the abundance of local maxima (peaks) and minima (dips) identified in the smoothed distribution of halos and
  dark matter (DM) on scales of $10-100$s Mpcs.
  The simulations include  Gaussian and local-type $\fnl$ non-Gaussian initial conditions. 
  The expression derived in the literature for the abundance (irrespective of height)  of peaks for Gaussian fields is surprisingly accurate for the evolved halo and DM density fields for
  all initial conditions considered.  Furthermore, the height distribution is very well fitted by a log-normal on quasi-linear scales.
  The abundance  as a function of scale depends on the cosmological parameters ($H_0$ and 
 background  matter densities) through the shape of the power spectrum,  but it is insensitive to the clustering amplitude. Further,  the abundance in the smoothed halo distribution is  substantially different in the  non-Gaussian from the  Gaussian simulations.   The  interpretation of this effect is straightforward in terms of 
  the scale dependence of halo bias in non-Gaussian models.
The abundance of extrema  extracted from three-dimensional large galaxy redshift surveys could be a  competitive 
probe of the cosmological parameters and  initial non-Gaussianity. It breaks the degeneracy between $\fnl$ and the clustering amplitude, making it  complementary to  counts of galaxy clusters and peaks in weak-lensing maps. 
\end{abstract}


\section{Introduction}
\label{sec:intro}

Peaks in the underlying mass density field are the most likely sites for the formation of halos where gas is expected to accrete and form galaxies
\citep{White1978}.
In the classical picture of \cite{PS}, matter in regions with  linear density contrast above a threshold $\delta_c$  is assigned to
halos of mass larger  than $M$,  where $M$ defines the smoothing of the density field. This implies that halos of mass $M$ form at
peaks with $\delta=\delta_c$ in the smoothed density contrast $\delta$.

Naturally, most studies have focused on peaks associated with halos. Indeed, 
statistical properties of local extrema \citep[e.g.][]{Adler1981} have gained a great deal of   attention in cosmology \citep{BBKS} (hereafter BBKS).
Correlations of halos and their distribution in relation to the  mass density field of the gravitationally dominant dark matter (DM), i.e. biasing
\citep[][]{Kaiser1984a}, have been studied
extensively with analytic methods and numerical simulations.
For Gaussian initial conditions and on sufficiently large scales,  halos follow a linear biasing relation, $\delta_\mathrm{h}=b\delta$ between the
halo number density contrast, $\delta_\mathrm{h}$ and the mass  density contrast $\delta$.
The bias factor $b$ depends on the height of the peaks associated with halos and on their mass. 
An important result obtained in simulations \citep{Dalal:2007cu}, and confirmed by analytic techniques
\citep{grinstein/wise:1986,Dalal:2007cu,Matarrese:2008nc,Slosar:2008hx}, is that  the presence
of  initial local-type non-Gaussianity introduces a peculiar scale dependence in the bias factor dubbed ``non-Gaussian bias''.
The specific form of $b(k)$ ($k$ is the wavenumber of a given scale) opens the window for probing initial non-Gaussianity based on the clustering
properties of galaxies in planned large  redshift surveys, e.g. Euclid  \citep{EuclidRB} and DESI \citep{DESICollaboration2016a}.

It is well known that non-Gaussianity strongly affects the tails of density probability distributions \citep{Adler1981,catelan/etal:1988}.
Several authors have further specialized these results to local density maxima of non-Gaussian density fields, where the non-Gaussianity is  either  of a
generic form 
\cite[e.g.][]{Catelan1988,Gay2012,codis/etal:2013,uhlemann/etal:2018} or developed via non-linear gravitational evolution of initial gaussian conditions
\citep[e.g.][]{Suginohara1991,matsubara:1994}. 
In particular, \cite{Gay2012,codis/etal:2013} considered the effect of a generic non-Gaussianity on extrema counts and Minkowski functionals of the dark
matter density field. 
In this work, we consider peaks and dips in cosmological density field smoothed on scales  much larger than those of galactic and galaxy cluster halos
($\ltsim 10$ Mpcs).
Using N-body simulations in large cosmological boxes, we  focus on the total number of local extrema for density fields constructed from the {\it halo}
distribution, as a proxy for a galaxy catalogue. Earlier analyses (\cite{Croft:1997rv,desoma,De:2009uz}) have used this type of statistics for constraining parameters related to the linear matter power spectrum on smaller scales ($\ltsim 10$ Mpcs).
Our goal is to assess the extent to which the abundance of extrema in three-dimensional (3D) fields inferred from current and forthcoming large galaxy  redshift surveys can be used as
a cosmological tool and, more specifically, a probe of local primordial non-Gaussianity.
As we shall see, our main findings have a straightforward interpretation in terms of the non-Gaussian bias.

We adopt standard notation. The mean total and baryonic mass densities (in units of the critical density) are denoted by 
$\Omega_\mathrm{m}$ and $\Omega_\mathrm{b}$, respectively. The Hubble constant is $H_0$ and $h=H_0/[100 \mathrm{\kms\; Mpc^{-1}} ]$.
The linear growth factor (normalized to unity at the present time) at redshift, $z$, is $D(z)$.
The outline of the paper is as follows. In \S\ref{sec:basics} we lay out   known relations between the number of extrema and the underlying power spectrum for Gaussian fields.
A description of the N-body simulations is provided in \S\ref{sec:simulations} and  the corresponding  results for the abundance of local extrema identified in
smoothed density fields derived from the DM and halo distributions are in  \S\ref{sec:results}. 
In \S\ref{sec:prospects}  we discuss the prospects for the application of the number of extrema as a test of cosmological parameters and conclude with a summary in 
\S\ref{sec:discussion}.

\section{Definitions and Theoretical Expectations}
\label{sec:basics}

We define  local maxima (peaks) in a smoothed random field, $f$,  as points in space where the spatial gradient is $\partial_\alpha f=0$ and the Hessian
$\partial_\alpha\partial_\beta f$ is negative  definite. Local minima (dips) are defined similarly  but with a positive definite Hessian. 
For a random Gaussian field, peaks and  dips have an equal total number \textit{per unit volume}, which was computed by BBKS to be  
\begin{equation}
n_{0}\approx 0.016 R_*^{-3}\; ,
\label{eq:npk}
\end{equation}
where 
\begin{equation}
R_*=\sqrt{3} \frac{\sigma_1}{\sigma_2}\; ,
\label{eq:rst}
\end{equation}
and  the spectral moments 
\begin{equation}
\sigma^2_j=\int \frac{k^2\dd k}{2\pi^2}P(k) W_R^2(k) k^{2j}\; .
\label{eq:sigi}
\end{equation} 
The expression for $n_0$ is independent of the clustering amplitude and it depends only on the shape of the power spectrum, $ P(k)$ of the field, and the smoothing
Kernel $W_R(k)$. 
For  $P(k)\sim k^n$, and a Gaussian smoothing window $W_R^2(k)=\exp(-k^2R^2)$, it is easy to see that, 
\begin{equation}
\frac{R}{R_*}=\left(\frac{n+5}{6}\right)^{1/2}\; .
\label{eq:Rn}
\end{equation}
The total number of peaks is  preserved under a local monotonous one-to-one mapping, $F(\delta)$,  of the density field. 
Thus we expect this quantity to be independent of time in the quasi-linear scales. On smaller scales, local extrema  tend to merge and diffuse,
leading to deviations from  expression Eq.~(\ref{eq:npk}) above.


In addition to the DM density field, we also examine peaks and dips in the smoothed distribution of halos. 
The corresponding spectral moments  $\sigma_j\equiv\sigma_{j,h}$ are given by
\begin{equation}
  \label{eq:sigmahi}
  \sigma_{j,h}^2 = \int\frac{k^2 dk}{2\pi^2}\, k^{2j} W_R^2(k) \Big[b^2(k) P(k) + \frac{1}{\bar n_h}\Big] \;. 
\end{equation}
The expression in square brackets is a model for the power spectrum of the halo distribution
where $P(k)$ here  refers to the  underlying density field and $b(k)$ describes the scale-dependent halo bias.
The term $1/ {\bar n_h}$ is due to the finite number of halos and approximated as a Poisson discreteness noise.
Using the simulations described below we have found that the added discreteness variance  is strongly suppressed for large smoothing and
is actually sub-Poissonian, in agreement with the findings of \cite{casas-miranda/etal:2002,Hamaus:2010im}.
On linear scales, the halo bias $b(k)$ is constant for Gaussian initial conditions but depends on the halo mass i.e. $b(k)= b^\mathrm{G}(M)$.

We also  consider local-type non-Gaussianity \citep{Salopek1990,Gangui,KomatsuSpergel} for which
the Bardeen potential $\Phi$ deep in matter domination is expanded around a random Gaussian field $\phi$ as
\begin{equation}
\Phi(\bx) = \phi(\bx)+f_{\rm NL} \left( \left[\phi(\bx)\right]^2-\langle \phi^2\rangle\right) \;. 
\end{equation}
The bispectrum of $\Phi$ induces the following scale dependence in the bias factor,
\begin{equation}
\label{eq:bNG}
  b(k) = b^\mathrm{G}(M) + \frac{\alpha(\fnl)}{k^2T(k)}\; ,
\end{equation}
where
\begin{equation}
  \alpha(\fnl) \equiv 3 \fnl \frac{\partial\ln\bar n_h}{\partial\ln\sigma_8} \frac{\Omega_m H_0^2}{D(z)c^2} \;, 
\end{equation}
and  $\bar n_h(M)$ is the   abundance of halos (per unit $M$) computed for the Gaussian field without the $\fnl$ terms.
When implementing Eq. \eqref{eq:sigmahi} to compare it to data (see Section \S\ref{sec:results}), we use the following approximation
\begin{equation}\label{eq:universal}
  \frac{\partial\ln\bar n_h}{\partial\ln\sigma_8} \approx \delta_c (b^{\rm G}(M)-1),
\end{equation}
with $\delta_c=1.687$, which is valid for universal mass functions and the spherical collapse model~\footnote{See \cite{Biagetti:2016ywx} for a
  quantitative analysis  on this approximation on  the same set of simulations, sim 1, used here.}.
We do not include  expressions \citep[e.g.][]{Gay2012} for the theoretical corrections  to  Eq.~(\ref{eq:npk}) due to $\fnl$ non-Gaussianity. Indeed, we will 
see below that the expression remains accurate provided that the appropriate $\sigma_i$ is used. 

\section{Simulations}
\label{sec:simulations}

Two sets of simulations, respectively in a $2\hgpc$ and a $3\hgpc$ box, are available for initial conditions generated from  $\Lambda$CDM initial power spectra
with slightly different cosmological parameters, as described in the Table. 
The simulations were run with the  Gadget2 \citep{Gadget2} N-body code on the Baobab cluster at the University of Geneva.
The initial particle displacements were implemented at $z_i=99$ using the public code 2LPTic \citep{Crocce2006} for realizations with Gaussian initial conditions
and its modified version \citep{Scoccimarro12} for non-Gaussian initial conditions of the local type. 
The transfer function for the smaller box (simulations 1, see Table) was obtained using the CLASS code \citep{Blas2011}.
This set contains runs for Gaussian initial conditions and two for local-type non-Gaussianity  respectively,  with $\fnl=250$ and $\fnl=-250$.
For each of these initial conditions, we obtain 8 random realizations  corresponding to different random seeds. 

The transfer function  of the second set, simulations 2, was  obtained using the CAMB code \citep{Lewis:1999bs}.
This set includes 3 types of  models: Gaussian initial conditions ($\fnl=0$) and non-Gaussian initial conditions, respectively, 
with $\fnl=100$ and $\fnl=-100$. For each type of models, we have 3 simulations corresponding to different random realizations of the initial conditions.
The Rockstar  \citep{Behroozi2013} algorithm is employed to identify halos, with linking length $\lambda=0.28$.

\begin{table}
   \centering
  \begin{tabular}{c c c c c c c}
    \hline \hline
 & $L$   & $N_\mathrm{p}$& $M_\mathrm{halo}$& $\sigma_8 $ & $\Omega_m$& $\Omega_b $ \\  
    \hline 
sim 1 & 2&  $1536^3$& $3.67$ & 0.85 & 0.3 & 0.0455 \\
sim 2& 3 &   $1024^3$& $37.9$ & 0.81 &0.272 & 0.0455  \\
 \hline         
\end{tabular}
  \caption{ Simulation parameters, where $L$ is the box size (in unit of $\hgpc$), $N_\mathrm{p}$ number of simulation particles,
    and $M_\mathrm{halo}$ is the  minimum halo mass identified in the simulation (in unit of $10^{12}\Msun$).
    Both, simulations 1\&2, include Gaussian and two choices for  non-Gaussian initial conditions. 
  Outputs of simulations 1
are available at $z=0$ and $z=1$, while only  the output at  $z=0$ is available for simulations 2.
In all simulations the Hubble parameter is $h=0.7$ and the spectral index of the initial power spectrum at large scales is $n_s=0.967$.}
\end{table}

\begin{figure}
 \includegraphics[width=0.45\textwidth]{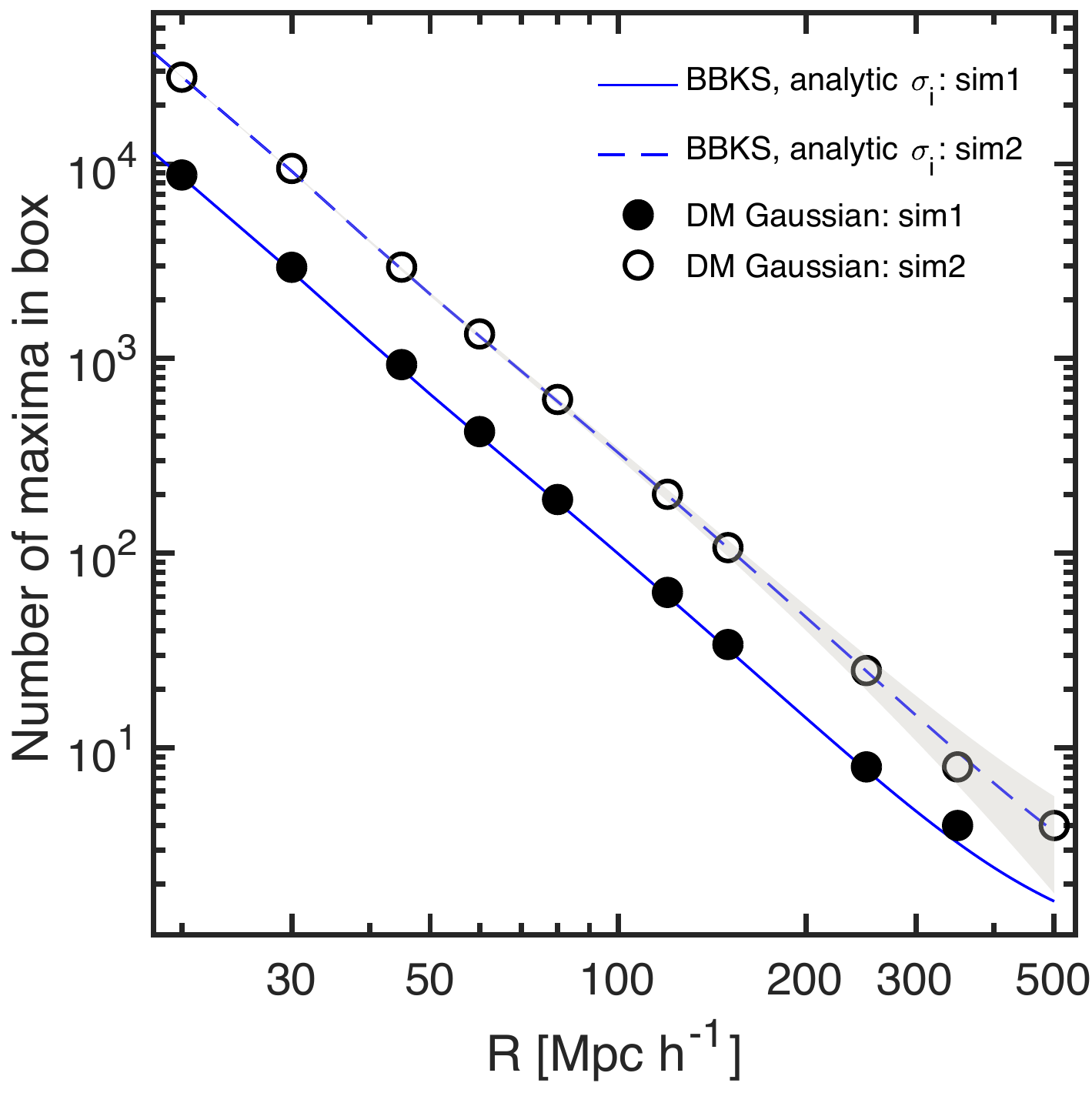} 
 \vskip 0.0in
 \caption{Total number of maxima versus the smoothing length, from   the  DM distribution in simulations 1 \& 2 for Gaussian initial conditions at redshift $z=0$
   The lines represent  the corresponding theoretical prediction using eq. \ref{eq:npk} and the shaded area 
 represents the $1\sigma$ shot-noise for 
 the larger simulation. }
\label{fig:box}
\end{figure}

Density fields are interpolated from the  DM and halo distributions in the simulation box 
 on a  $512^3$ cubic grid   using the Clouds-in-Cells (CIC) scheme.  
 The grid spacing is thus $3.9\hmpc $
and $5.85\hmpc$, for simulations 1 \& 2, respectively.  The density fields were additionally  smoothed with a Gaussian window 
of 8 different widths in the range $20\hmpc$ to $500\hmpc$. For each smoothed field, local maxima (minima) were identified as grid points surrounded by 
grid points with lower (higher) density values.
Fig.~\ref{fig:box} shows the total  number of maxima 
in the smoothed DM density field in the full boxes of  simulations 1 \& 2 at $z=0$. 
The theoretical predictions obtained from the BBKS expression (Eqs.~\ref{eq:npk}-\ref{eq:rst}) using the linear power spectrum $P(k)=P_\mathrm{L}(k)$ 
 for the two models are also shown, as indicated in the figure\footnote{In performing the integration in Eq.~(\ref{eq:sigi}), it is important to impose a low $k$ cutoff 
corresponding to the finite box size of the simulations. }.
The shaded area encompasses the  expected range of (1$\sigma$) shot-noise for simulations 2. 
The number drops like $R^{-3}$, consistently with Eq.~(\ref{eq:npk}) since $R_*\propto R$ upto a factor of $\mathcal{O}(1)$ which depends on the shape of the power
spectrum at scale $R$ (cf. Eq.~\ref{eq:Rn}).  
The  figure refers to the Gaussian simulations only. A similar figure can be found in \citep{2011MNRAS.413.1961L},
but for comparison of the theoretical expression with 
peaks identified in the initial conditions of  their simulations.

\section{Results}
\label{sec:results}

\subsection{Total number of minima and maxima}

Gaussian initial conditions imply equal probability of producing peaks and dips, up-to fluctuations due to the finite box size.
However, on scales  $\ltsim  10$s of Mpcs, non-linear gravitational evolution breaks the initial symmetry through
the merging and smearing of dips and peaks.
For  non-Gaussian initial conditions, the   statistical symmetry between  maxima and minima is already broken 
initially.

We choose to first analyze  the (total) number $n_\mathrm{1}$, per unit volume, of minima, $n_\mathrm{min}$, and maxima,
$n_\mathrm{max}$ in the simulations. The differences between the abundance of minima and maxima will be discussed at a later stage.
More precisely, we consider
\begin{equation}
\label{eq:n1}
n_\mathrm{1} =\frac{1}{2}\big(n_\mathrm{min}+n_\mathrm{max}\big) \; , 
\end{equation}
which is computed from the smoothed density fields for the various simulations. An  advantage of $n_1$ is that it boosts  the statistical significance of the measured abundance. For a  Gaussian field, $n_1=n_0$ given in Eq.~(\ref{eq:npk}).
Inclusion of non-Gaussian terms modify the abundance of either the minima or maxima by a leading-order
correction proportional to the skewness of the density field and its derivatives \citep{Gay2012}.
The combined leading order correction for both minima and maxima cancel out in the expression of $n_\mathrm{1}$.
Consequently, the BBKS prediction Eq.~(\ref{eq:npk}) remains valid up to a small correction of order $\fnl^2$.

\begin{figure}
 \includegraphics[height=1.2\textwidth]{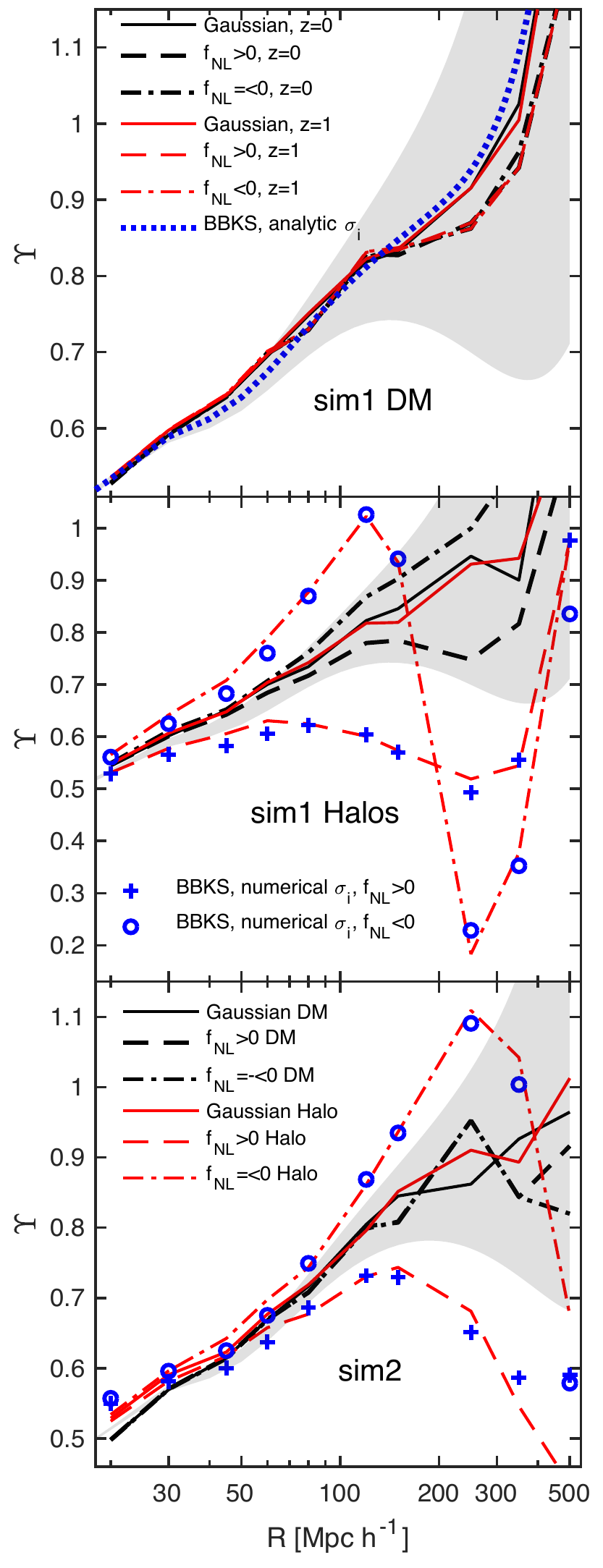} 
 \caption{The quantity $\Upsilon$ as estimated from Eq.~(\ref{eq:Rfromn}).
   \textit{Top:} from the number of peaks and throughs in the dark matter distribution of simulations 1.
   \textit{Middle: } The same the {\it Top}, but for the halo distribution.
   \textit{Bottom:} For DM and halos for simulations 2, at $z=0$ only.
   In all panels, the grey area represents the shot-noise estimated from the expression with using the theoretical linear power spectrum.}
\label{fig:panels}
\end{figure}

According to Fig.~\ref{fig:box}, differences in $n_1$ between the simulations are visually hard to examine directly. 
Thus,  we consider the statistic,  
\begin{equation}
\label{eq:Rfromn}
\Upsilon\equiv \frac{n_1R^3}{0.016}
\end{equation}
where $R$ is the width of the smoothing window. According to Eq.~(\ref{eq:npk}), for a Gaussian field   $\Upsilon=({R}/{R_*})^3$.
The three-panel Fig.~\ref{fig:panels} summarizes the main results. 
The top panel   plots $\Upsilon$, averaged over the 8 random realizations in simulations 1,  against the smoothing length, $R$, for the DM density field.
The shaded area represents the 1$\sigma$ shot-noise in $n_1$  corresponding to the finite number of 
peaks and dips in the simulation box. 
It is estimated as $\sqrt{n_0 L^3/2}$ where  $n_0$ is the theoretical 
value  according to  Eq.~(\ref{eq:npk}) and the factor of $1/2$ arises from the definition of $n_1$ which involves both minima and maxima.
We have checked that the scatter from the  8 individual runs (not shown for clarity) is consistent with this estimate of the shot-noise.
For our  Gaussian as well as  non-Gaussian simulations, the results in the top panel  
for $z=1$ and $z=0$ are almost identical.
The dotted line  shows   $(R/R_*)^3$ computed according to the theoretical expression  Eq.~(\ref{eq:rst}) derived for Gaussian fields, where $\sigma_i$ are computed
using Eq.~(\ref{eq:sigi}) with the  initial power spectrum $P_\mathrm{L}(k)$.
There is a reasonable  match between the dotted curve  and $(R/R_*)^3$ derived from $n_1$ for the Gaussian simulations (black and red solid curves).
Overall, the impact of $\fnl$ is very small, in agreement with the fact that, for dark matter, $n_1$ depends on $\fnl$ only at order $\fnl^2$.

The middle panel refers to  results  obtained from the halo distribution in  simulations 1.
The  solid curves corresponding to  the Gaussian simulations   at $z=1$ and $z=0$  are similar.
In great contrast to the upper panel, both $\fnl=250$ and $\fnl=-250$ models (dashed and dash-dotted lines) at the two redshifts are substantially
different.
It is interesting to check how well the BBKS expression in  Eq.~(\ref{eq:rst}) fits the $\Upsilon$  computed from the halos in the non-Gaussian simulations. 
To do that we compute  $(R/R_*)^3$ using Eq.~(\ref{eq:rst}) for $\sigma_1$ and $\sigma_2$ computed directly from the halo
density fields.
The results are plotted as the plus signs and circles, respectively, for the $\fnl=250$ and $\fnl=-250$ simulations.
We present the  $z=1$ case only but the excellent agreement of $\Upsilon$ with $(R/R_*)^3$  computed from $n_\mathrm{1}$ also holds at $z=0$ .
The bottom panel summarizes results for simulations 2 ($z=0$) of the larger box.
The halos in these simulations have a larger mass and therefore follow a different biasing relation than halos in simulations 1, yielding different
quantitative results.
For these simulations also, the BBKS expression (computed with $\sigma_i$ measured in the simulations), shown as the plus signs and circles, furnishes an
excellent match.
Therefore,  despite the fact that relations Eqs.~(\ref{eq:npk}) and (\ref{eq:rst}) are formally obtained for Gaussian fields, 
they remain accurate for the non-Gaussian fields considered here, provided the actual  $\sigma_i$ are used. 

\begin{figure}
 \includegraphics[width=0.45\textwidth]{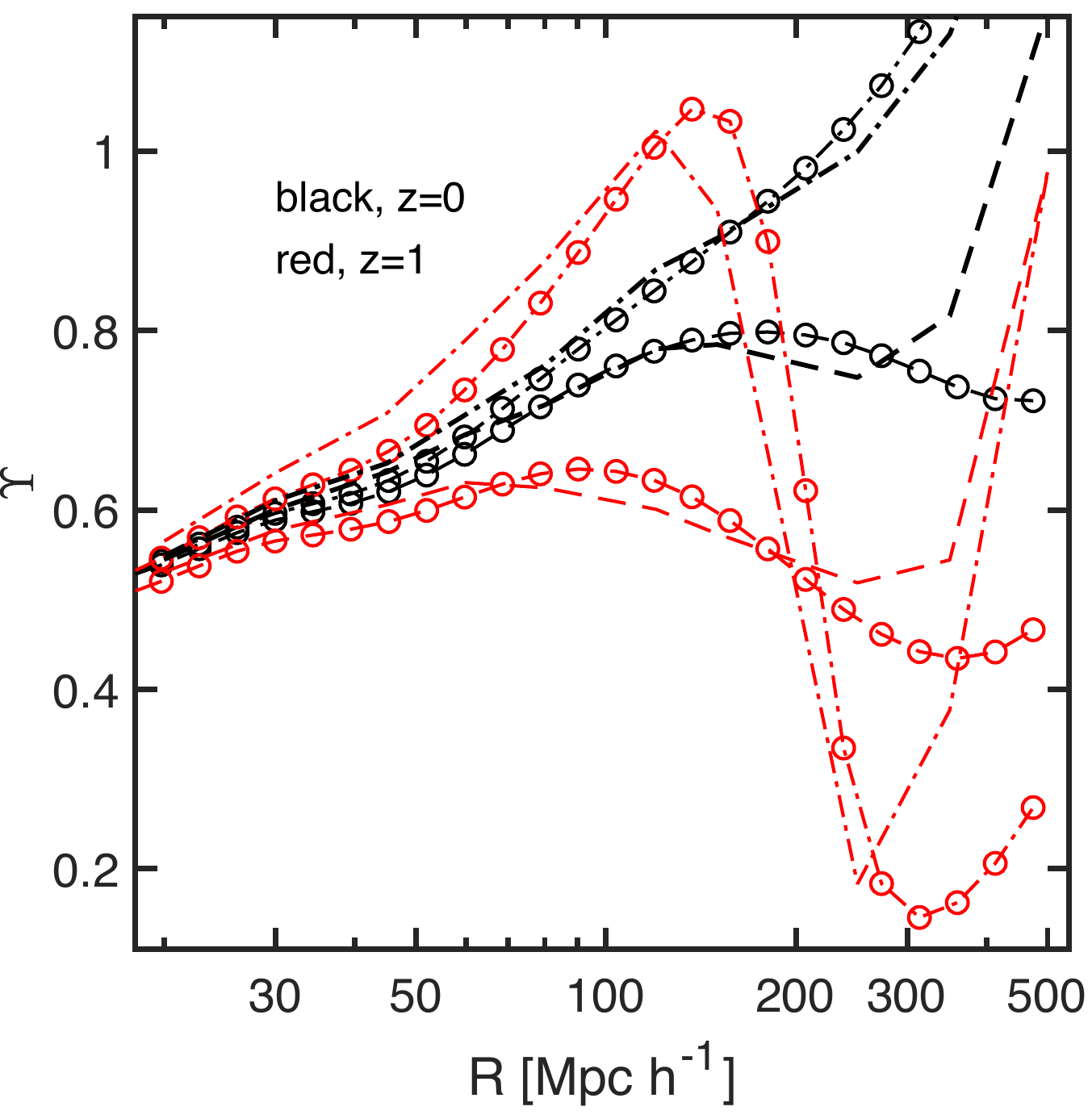} 
 \caption{A test of the analytic prediction  for the non-Gaussian model. The dashed and dash-dotted curves are   taken from the middle panel in the previous figure. 
 The curves with the circles plot $(R/R_*)^3$ computed with the approximate $\sigma_i$ given in Eqs.~(\ref{eq:sigmahi}-\ref{eq:bNG}) }
\label{fig:fnltheory}
\end{figure}

In Fig.~\ref{fig:fnltheory} we compare the theoretical expectation of Eq. \eqref{eq:sigmahi} against  $\Upsilon $ measured from the non-Gaussian simulations. 
The theoretical curve fits good the data on scales $R \lesssim 100$Mpc/h and provides a qualitatively good description at all scales.
Deviations may be due in part to our approximation Eq. \eqref{eq:universal} and, especially at large scales, to the finite box size of the simulations.

To conclude this Section, we note that the effect of $\fnl$ on $\Upsilon$ is only weakly degenerate with that of $\sigma_8$ because $n_1$ primarily depends on the
ratio of spectral moments $\sigma_{1,h}/\sigma_{2,h}$.

\subsection{Asymmetry and height distribution}

\begin{figure} 
 \includegraphics[width=0.45\textwidth]{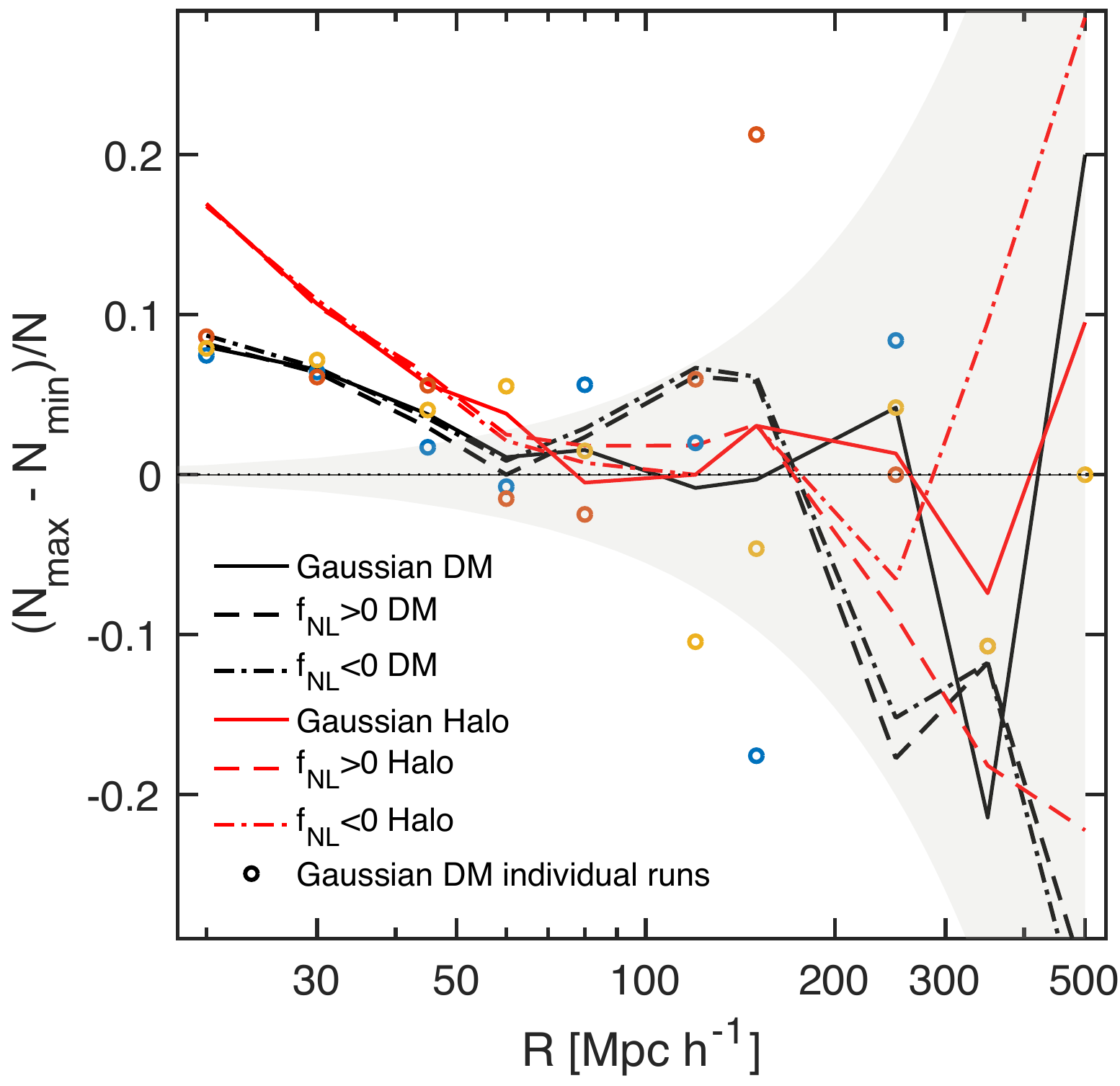}
 \caption{ The relative difference between the total number of maxima and minima in simulations 2, versus the smoothing width, at redshift $z=0$.}
\label{fig:asymm}
\end{figure}

So far we have considered $n_1$, without distinguishing between minima and maxima. 
In Fig.~\ref{fig:asymm}, we examine the asymmetry between the abundances of  minima and maxima as a function of the smoothing width for simulations 2 at redshift $z=0$.
There is a clear excess  of $N_\mathrm{max}$, which is significantly above the level of the shot-noise (grey area). 
The trend is reversed at larger scales for both DM non-Gaussian models, but it becomes immersed in the shot-noise.
Results of the three individual runs for the Gaussian DM simulation are also shown.
It is clear that the shot-noise estimated theoretically as described above (grey area) is consistent with the scatter in the individual runs.

\begin{figure} 
 \includegraphics[width=0.45\textwidth]{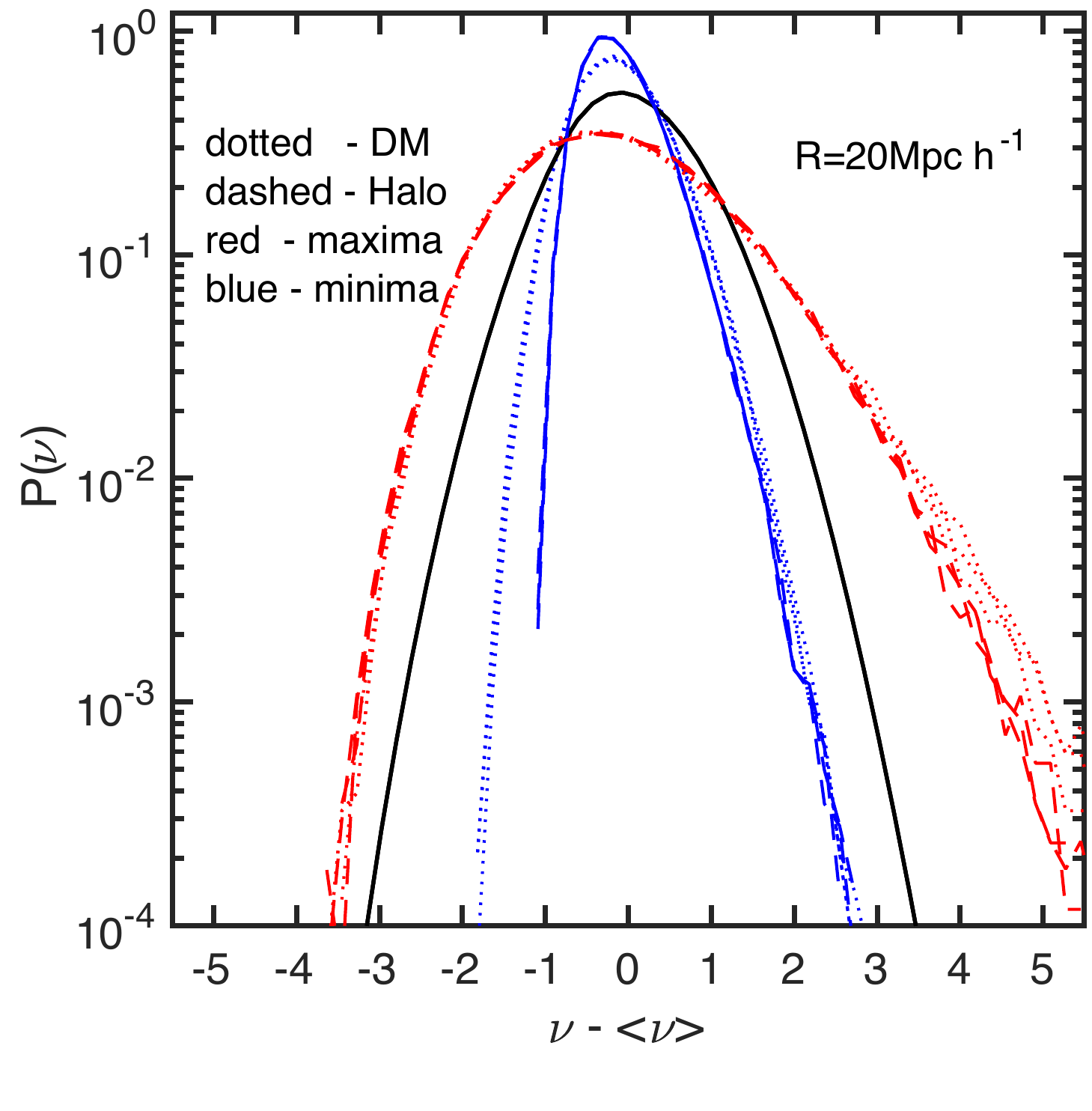}
 \includegraphics[width=0.45\textwidth]{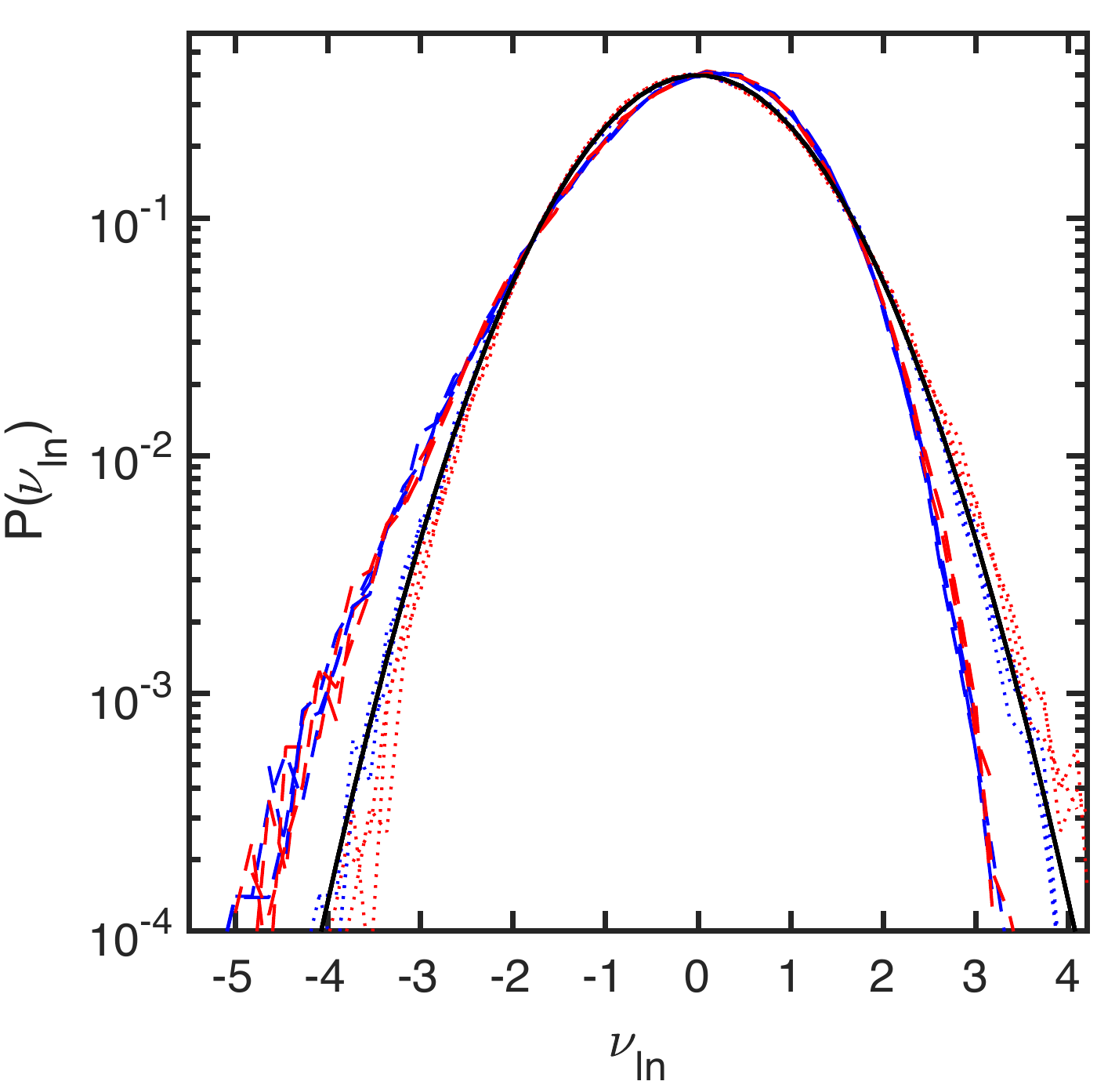} 
 \caption{ \textit{Top:} The PDF of $\nu$ at minima and maxima in simulations 2, as indicated in the figure.
 The black curve is the theoretical prediction for $P(\nu)$ given in BBKS.
 \textit{Bottom:} The PDF of $\nu_\mathrm{ln}$ at minima and maxima for simulations 2. Here
 the black line is a Gaussian with zero mean and unit variance. }
\label{fig:PDFnu}
\end{figure}

\begin{figure} 
 \includegraphics[width=0.45\textwidth]{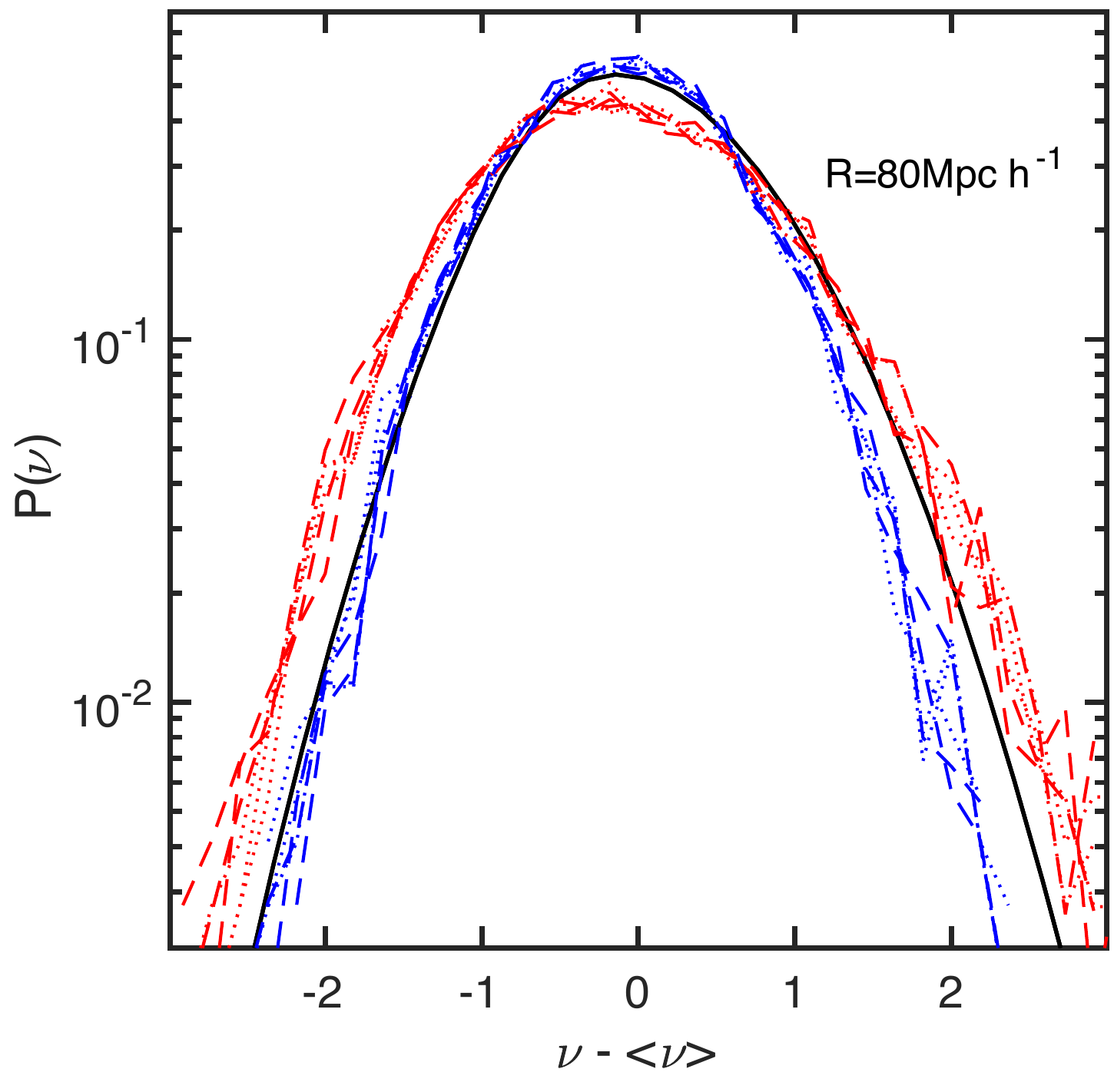}
 \includegraphics[width=0.45\textwidth]{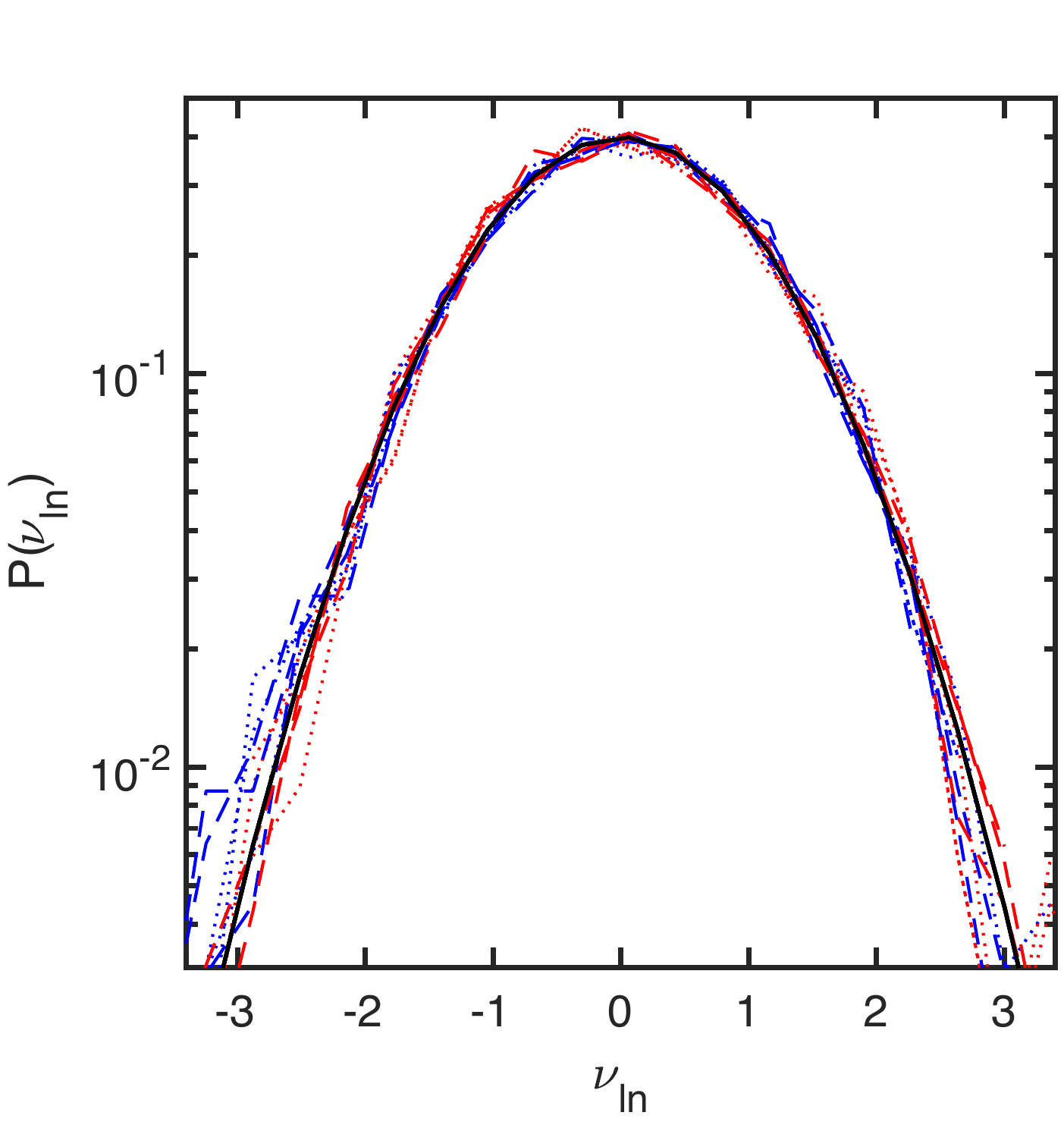} 
 \caption{ The same as  the previous figure, but for $R=80\hmpc$. }
\label{fig:PDFnu80}
\end{figure}

We explore the  probability density distribution (PDF) of the value of the densities at the minima and maxima.
We  define, 
\begin{equation}
\nu=\frac{\delta}{\sigma_0}\quad {\rm and}  \quad \nu_\mathrm{ln}=\frac{\ln(1+\delta)-\mu}{\sigma_\mathrm{ln}}
\end{equation}
where $\sigma_0$ is the rms of density field all over space while 
 $\mu $ and $\sigma_\mathrm{ln}$ are the mean and rms of the values of $\ln(1+\delta)$ at  either the minima or maxima.
 The quantity  $\nu_\mathrm{ln}$ is motivated by the result that the PDF of the density field is well approximated by a log-normal distribution \citep[e.g.][]{Coles91,Kofman}.
In Fig.~\ref{fig:PDFnu} and \ref{fig:PDFnu80}   we plot the PDF of  $\nu $ (top) and 
$\nu_\mathrm{ln}$ (bottom) for a smoothing of $R=20\hmpc$ and $80\hmpc$ for simulations 2. 
The 3 curves of  each line-style correspond to the Gaussian and 2  non-Gaussian simulations.
It is evident that the  
PDF of densities at either maxima or minima  is weakly sensitive to whether the initial conditions were Gaussian or not.
This is expected given that corrections arise  at order $f_{\rm NL}^2$ as noted above.
Thus, for clarity, the plot does not indicate which of the simulations is shown. 
For $R=20\hmpc$, the BBKS theoretical prediction for $P(\nu)$  (expression 4.3 in their paper) shown as the black in the top panel, is a poor fit to any of the  PDFs  measured in the simulations.
However, $P(\nu_\mathrm{ln}) $ for the DM density field (dotted), exhibit  only minor differences at the tails, where  the PDF for maxima is slightly skewed to positive values
relative to the Gaussian (black in the bottom panel), the distribution at minima is negatively skewed. The rather small differences between 
the PDF from the halos   and the corresponding DM are due to deviations from linear  biasing. 
Fig.~\ref{fig:PDFnu80} shows the same results, but for $R=80\hmpc$. This large smoothing greatly reduces the effect of non-linear evolution, bringing 
the BBKS theoretical  PDF (black curve, top panels)  closer 
to the measured PDF than  it is for   $R=20\hmpc$. 
The log-normal curve (black, bottom) remains a good fit to $P(\nu_\mathrm{ln} $ for the DM although   not as good as in the smaller smoothing.  It is interesting that the log-normal describes the halo PDF fairly well for this smoothing. 
At $R=20\hmpc$ and $80\hmpc$  the halo bias in the Gaussian and non-Gaussian simulations are  small (cf. Eq.~(\ref{eq:bNG}). This  explains the similarity between the halo PDFs in the simulations irrespective of the initial statistic. 

\section{Abundance of extrema as  a cosmological test}
\label{sec:prospects} 

We offer a preliminary assessment of using total number of peaks and dips as a test of cosmological models.
A proper analysis should take into account the covariance between the abundances corresponding to different smoothing scales.
However, this task is beyond the scope of the current paper. 
Instead, we will focus on the expected discriminatory power of extrema abundance at distinct scales. 
As an example, we consider the {\small Euclid} mission \citep{EuclidRB}, which will target emision line galaxies in
the redfshift range $0.9<z<1.8$ across $\sim 35\%$ of the sky. 
For Planck's cosmological parameters, the corresponding survey   volume is $48 (\hgpc)^3$. Furthermore,
the typical host halo mass is $\sim 10^{11-12}\Msun$, in broad agreement with the minimum halo mass resolved
in simulations 1.

We wish to  assess  the ability that  a  measured total number $N$ of
extrema  in a survey can reject a certain model given the hypothesis of an assumed fiducial   underlying model. 
For this purpose, we assume that $N$ follows a Poisson distribution 
\begin{equation}
\label{eq:Poisson}
P_{\bar N}(N)= \frac{{\bar N}^N}{N!} \mathrm{e}^{-\bar N}\; ,
\end{equation} 
where $\bar N$ is the mean  number expected in a particular given model. 
Given an observed $N$, 
the preferred of two competing models with expected mean numbers ${\bar N}_1$ and ${\bar N}_2$, respectively, is determined by 
\begin{eqnarray}
\label{eq:D}
D_{_{\bar N_1 \bar N_2}}&=&-2\ln \frac{P_{{\bar N}_1}}{P_{{\bar N}_2}}\nonumber \\
&=& 2 N\ln\frac{\bar N_2}{\bar N_1}+2({\bar N_1-\bar N_2})\; .
\end{eqnarray}
The mean value of $D$ over all measurements, which we loosely denote by $\Delta \chi^2$ is 
\begin{eqnarray}
\label{eq:chis}
\Delta \chi^2&=&\sum_N P_{\bar N} D_{_{\bar N_1 \bar N_2}} \nonumber \\
&=&2\bar N \ln\frac{\bar N_2}{\bar N_1}+2({\bar N_1-\bar N_2})\; ,
\end{eqnarray}
where we have used $\sum_N P_{\bar N}(N)=1$ and $\sum_N N P_{\bar N}(N)=\bar N$.
For  $\bar N_2=\bar N$, the quantity $\Delta \chi^2$  yields the confidence level with which 
a model with $\bar N_1$ can be rejected if the underlying model is $\bar N$.
We use this statistic  to assess whether  the abundance of dips and peaks 
can be used to reject certain models given a Gaussian cosmological model with fiducial cosmological parameter. 
We focus on $\Omega_m$ and $\fnl$, separately.  

\begin{figure} 
 \includegraphics[width=0.45\textwidth]{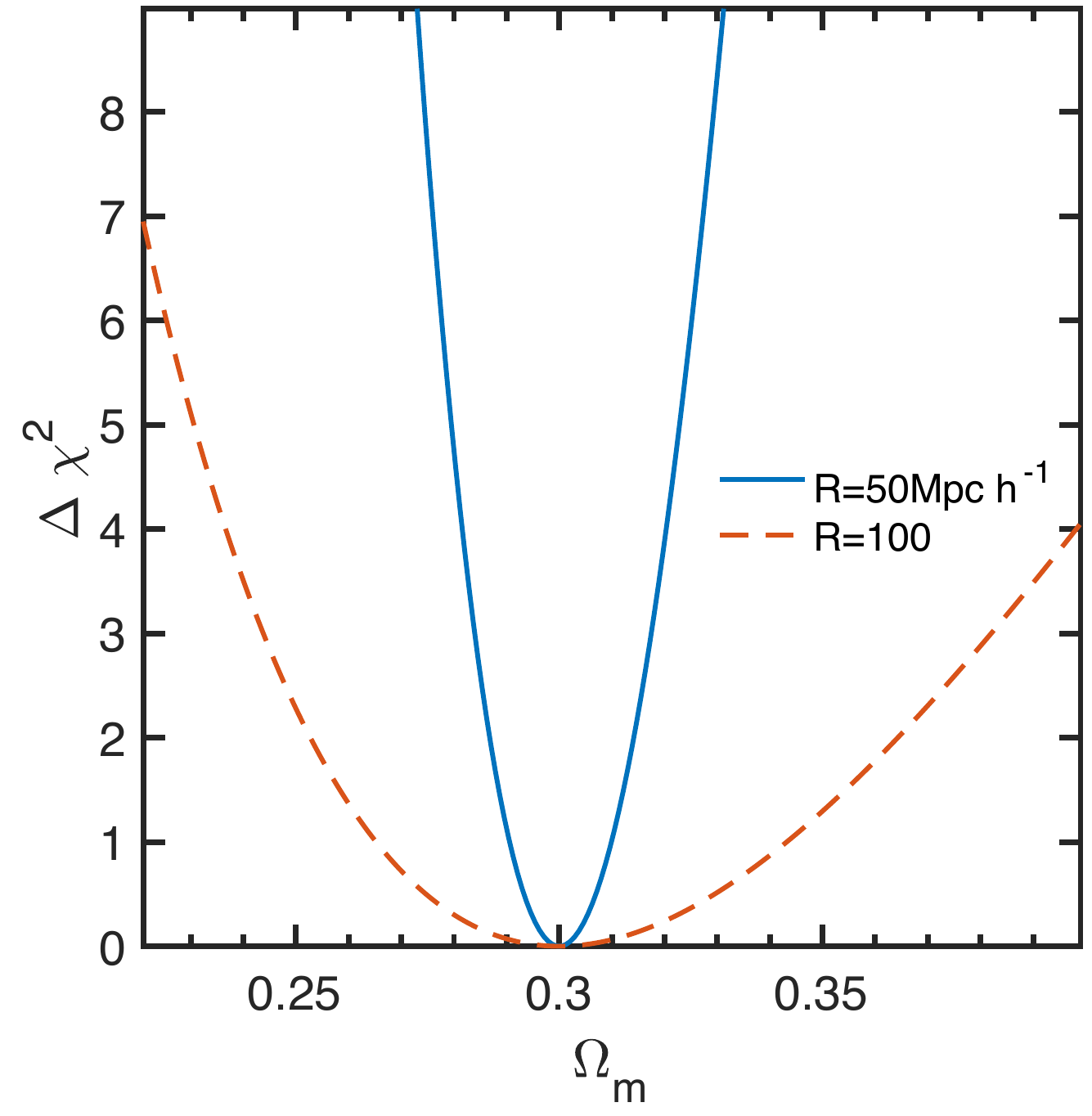}
 \caption{ Abundance of dips and peaks as a cosmological test for estimating $\Omega_\mathrm{m}$ from a  survey like {\small  Euclid}. 
 Values of $\Delta \chi^2=1$ correspond to $1\sigma$ limits from the fiducial value of $\Omega_\mathrm{m}$. }
\label{fig:chiomega}
\end{figure}

Fig.~\ref{fig:chiomega} examines  $\Delta \chi^2$ as a function of the matter density $\Omega_\mathrm{m}$.
Here,  $\bar N $ is computed using  Eq.~(\ref{eq:npk}-\ref{eq:sigi}) for fiducial DM linear 
power spectrum with the cosmological parameters corresponding to simulations 1.
The same parameters with the exception of $\Omega_\mathrm{m}$ are used in the same expression to derive $\bar N_1 $.
This figure, therefore,  refers to a Gaussian model ($\fnl=0$) and, in addition to DM density fields,
it is also relevant for halos with linear constant bias with respect to the DM. 
Only two filtering scales are considered, as indicated in the figure. 
It is remarkable that for $R=50\hmpc$ the $1\sigma $ level ($\Delta \chi^2=1$) is at $\Delta \Omega\approx\pm 0.01$ from the fiducial $\Omega_m=0.3$.
It should be pointed out that for the $\Lambda$CDM linear power spectrum the abundance on a filtering scale given in $\hmpc$ is degenerate with respect
to $\Omega_m h$ and $\Omega_b h$.
This sensitivity to $\Omega_m$ declines  rapidly at $R=100\hmpc$ due to the $1/R^3$ dependence of the number of dips and peaks. 

\begin{figure} 
 \includegraphics[width=0.45\textwidth]{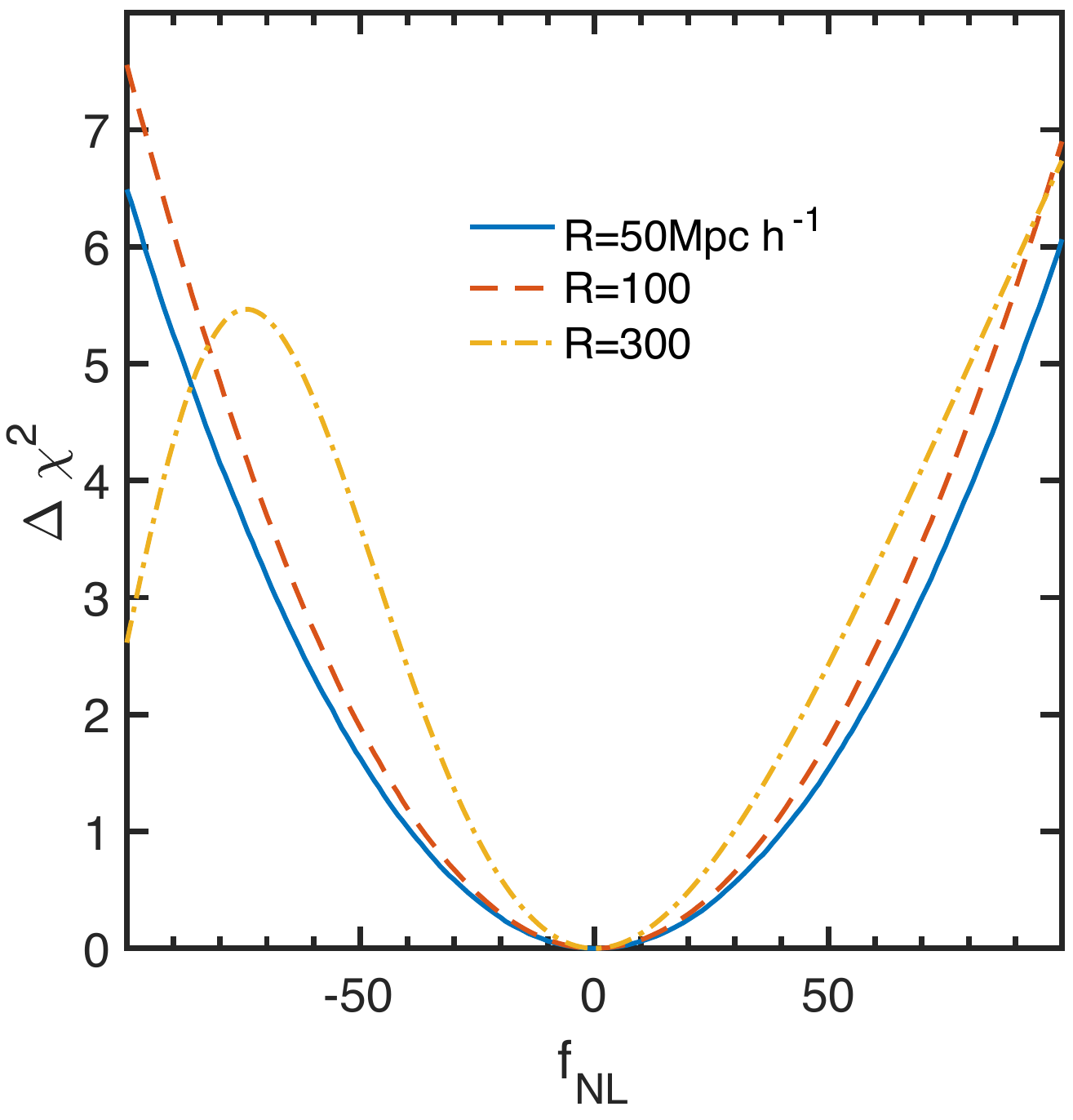}
 \caption{  The same as the previous figure but for $\fnl$ instead of $\Omega_\mathrm{m}$. }
\label{fig:chifnl}
\end{figure}

The sensitivity to $\fnl$ is demonstrated in Fig.~\ref{fig:chifnl}  plotting $\Delta \chi^2$ with $\bar N$  from the fiducial model and 
$\bar N_1$ for $\fnl\ne 0$  but with all other parameters fixed at the fiducial values. These curves refer to filtered halo distribution 
where the theoretical expressions in 
Eqs.~(\ref{eq:sigmahi}-\ref{eq:bNG}) are used in Eq.~(\ref{eq:npk})  to derive the mean number of dips and peaks $\bar N$ in a Euclid
volume survey at $z=1$.
In these calculations, we consider a halo mass  distribution consistent with simulations 1, with a minimum mass of  $3.67\times 10^{12}h^{-1} M_\odot$.
For this mass threshold, we have seen in the previous section that   the theoretical
predictions are in reasonable agreement with the simulations.
The sensitivity to $\fnl$ is improved for the larger filtering widths, $R$ thanks to the stronger $\fnl$-dependence of halo bias 
on larger scales. For $R=300\hmpc$, we find $\Delta \chi^2=1$ for deviations $\Delta \fnl\approx \pm 25 $.
This is encouraging especially if combined with measurements as a function of filtering scales and different halo masses.

\section{Discussion and Conclusions}
\label{sec:discussion}

Locating points of maxima and minima is straightforward even for 3D density fields estimated from   realistic galaxy redshift surveys.
Since the total abundance is computed irrespective of  height, 
it  should be robust against the  details of how the density field is estimated from the data. 
The total abundance is also insensitive to  redshift space distortions,
which in any case  can be modeled with standard perturbation theory for smoothing widths $R\gtrsim 50\hmpc$ \citep{codis/etal:2013}
\citep[see also][]{lam/etal:2010}.
Further, it  depends explicitly  only  on the shape of the power spectrum. 
 Any dependence on the amplitude (e.g. $\sigma_8$)  is indirectly  encoded 
in non-linear corrections to the shape of the gravitationally evolved power spectrum.
The lack of sensitivity of this abundance statistics on the amplitude thereby implies  
that it breaks most of the degeneracy
between $\fnl$ and the primordial amplitude of scalar perturbations, which  arises in measurements of 
 galaxy clusters counts and  shear peaks in weak lensing maps for instance
\citep[this degeneracy can also be broken by combining clusters and voids, see][]{kamionkowski:voids}.
We have demonstrated  that a primordial non-Gaussianity of the local-$\fnl$ type imprints a strong signal in the abundance of peaks
and dips of the halo density field owing to the non-Gaussian bias.
An important result of the current paper is that the BBKS prediction derived for Gaussian density field can account for this effect reasonably well,
provided that the matter power spectrum is replaced by the halo power spectrum.
Therefore, the abundance of peaks and dips (a 1-point statistics) is sensitive to the scale-dependent bias in the halo power spectrum (a
2-point statistics), like the covariance of cluster counts \citep{cunha/etal:2010}.
This effect disappears when the density field perfectly traces the matter distribution as is the case for shear peaks for instance.

We have made a preliminary assessment of the  applicability of  the total abundance  statistics as a test of $\fnl$ for a survey with specifications similar to those of the {\small Euclid}
mission \citep{EuclidRB}.
From a measurement at a single smoothing scale $R$, we obtain an uncertainty of $\Delta\fnl=25$ (for $R=300\hmpc$) and 40
(for $R=50\hmpc$).
This suggests that a measurement combining different smoothing scales and halo masses should be able to achieve a sensitivity of
$\Delta\fnl\lesssim 10$. While the sensitivity of this approach will likely be worse than the limits set by the latest CMB measurements
from Planck, $\fnl=0.8\pm 5$ \citep{PlanckPNG}, this approach should be competitive with galaxy clusters and shear peak counts in
weak-lensing maps,
for which the forecasted uncertainty is $\Delta\fnl\sim 9$ \citep[e.g.][for a galaxy survey like {\small eROSITA}]{pillepich/etal:2012}
and $\Delta\fnl\sim 13$ \citep[e.g.,][for a weak-lensing survey with Euclid specifications]{marian/etal:2011}, respectively.
However, our approach may also be affected by the Eddington bias that plagues galaxy cluster counts or shear peaks. Namely, additive
noise in the data will presumably increase the number of peaks while reducing the number of dips, which would mimic a small positive
$\fnl$. We will defer a more detailed study of this effect to future work. 

The abundance of extrema  depends on  the cosmological parameters of the background cosmology. Here we explored the dependence on  $\Omega_m$ alone with very encouraging results of an accuracy at the level of $\Delta \Omega_\mathrm{m}\sim 0.01$. 
For a given filtering scale given in $\hmpc$, the abundance  depends is nearly degenerate with the combination $\Omega_\mathrm{m}h$. Thus,  this result regarding   $\Omega_\mathrm{m}$ could alternatively by expressed as an accuracy of 
$0.7\kms$ on  $H_0$ if all other parameters are fixed.

\section*{Acknowledgements}

This research was supported by the I-CORE Program of the Planning and Budgeting Committee,
THE ISRAEL SCIENCE FOUNDATION (grants No. 1829/12 and No. 203/09 for AN; No. 1395/16 for VD)
and the Asher Space Research Institute.
M.B. acknowledges support from Delta ITP consortium, a program of the Netherlands Organisation for Scientific Research (NWO)
that is funded by the Dutch Ministry of Education, Culture and Science (OCW).

\bibliographystyle{mnras}

  \bibliography{references}

\begin{thebibliography}{}
\makeatletter
\relax
\def\mn@urlcharsother{\let\do\@makeother \do\$\do\&\do\#\do\^\do\_\do\%\do\~}
\def\mn@doi{\begingroup\mn@urlcharsother \@ifnextchar [ {\mn@doi@}
  {\mn@doi@[]}}
\def\mn@doi@[#1]#2{\def\@tempa{#1}\ifx\@tempa\@empty \href
  {http://dx.doi.org/#2} {doi:#2}\else \href {http://dx.doi.org/#2} {#1}\fi
  \endgroup}
\def\mn@eprint#1#2{\mn@eprint@#1:#2::\@nil}
\def\mn@eprint@arXiv#1{\href {http://arxiv.org/abs/#1} {{\tt arXiv:#1}}}
\def\mn@eprint@dblp#1{\href {http://dblp.uni-trier.de/rec/bibtex/#1.xml}
  {dblp:#1}}
\def\mn@eprint@#1:#2:#3:#4\@nil{\def\@tempa {#1}\def\@tempb {#2}\def\@tempc
  {#3}\ifx \@tempc \@empty \let \@tempc \@tempb \let \@tempb \@tempa \fi \ifx
  \@tempb \@empty \def\@tempb {arXiv}\fi \@ifundefined
  {mn@eprint@\@tempb}{\@tempb:\@tempc}{\expandafter \expandafter \csname
  mn@eprint@\@tempb\endcsname \expandafter{\@tempc}}}

\bibitem[\protect\citeauthoryear{Adler}{Adler}{1981}]{Adler1981}
Adler R.~J.,  1981, {The Geometry of Random Fields}.
Chichester: Wiley, \mn@doi{10.1137/1.9780898718980}

\bibitem[\protect\citeauthoryear{Bardeen, Bond, Kaiser  \& Szalay}{Bardeen
  et~al.}{1986}]{BBKS}
Bardeen J.~M.,  Bond J.~R.,  Kaiser N.,   Szalay A.~S.,  1986, \mn@doi [ApJ]
  {10.1086/164143}, 304, 15

\bibitem[\protect\citeauthoryear{Behroozi, Wechsler  \& Wu}{Behroozi
  et~al.}{2013}]{Behroozi2013}
Behroozi P.~S.,  Wechsler R.~H.,   Wu H.~Y.,  2013, \mn@doi [ApJ]
  {10.1088/0004-637X/762/2/109}, 762

\bibitem[\protect\citeauthoryear{Biagetti, Lazeyras, Baldauf, Desjacques  \&
  Schmidt}{Biagetti et~al.}{2017}]{Biagetti:2016ywx}
Biagetti M.,  Lazeyras T.,  Baldauf T.,  Desjacques V.,   Schmidt F.,  2017,
  \mn@doi [Mon. Not. Roy. Astron. Soc.] {10.1093/mnras/stx714}, 468, 3277

\bibitem[\protect\citeauthoryear{Blas, Lesgourgues  \& Tram}{Blas
  et~al.}{2011}]{Blas2011}
Blas D.,  Lesgourgues J.,   Tram T.,  2011, \mn@doi [JCAP]
  {10.1088/1475-7516/2011/07/034}, 2011

\bibitem[\protect\citeauthoryear{{Casas-Miranda}, {Mo}, {Sheth}  \&
  {Boerner}}{{Casas-Miranda} et~al.}{2002}]{casas-miranda/etal:2002}
{Casas-Miranda} R.,  {Mo} H.~J.,  {Sheth} R.~K.,   {Boerner} G.,  2002, \mn@doi
  [\mnras] {10.1046/j.1365-8711.2002.05378.x}, \href
  {http://adsabs.harvard.edu/abs/2002MNRAS.333..730C} {333, 730}

\bibitem[\protect\citeauthoryear{{Catelan}, {Lucchin}  \&
  {Matarrese}}{{Catelan} et~al.}{1988a}]{catelan/etal:1988}
{Catelan} P.,  {Lucchin} F.,   {Matarrese} S.,  1988a, \mn@doi [Physical Review
  Letters] {10.1103/PhysRevLett.61.267}, \href
  {http://adsabs.harvard.edu/abs/1988PhRvL..61..267C} {61, 267}

\bibitem[\protect\citeauthoryear{Catelan, Lucchin  \& Matarrese}{Catelan
  et~al.}{1988b}]{Catelan1988}
Catelan P.,  Lucchin F.,   Matarrese S.,  1988b, \mn@doi [PRL]
  {10.1103/PhysRevLett.61.267}, 61, 267

\bibitem[\protect\citeauthoryear{{Codis}, {Pichon}, {Pogosyan}, {Bernardeau}
  \& {Matsubara}}{{Codis} et~al.}{2013}]{codis/etal:2013}
{Codis} S.,  {Pichon} C.,  {Pogosyan} D.,  {Bernardeau} F.,   {Matsubara} T.,
  2013, \mn@doi [\mnras] {10.1093/mnras/stt1316}, \href
  {http://adsabs.harvard.edu/abs/2013MNRAS.435..531C} {435, 531}

\bibitem[\protect\citeauthoryear{Coles \& Jones}{Coles \&
  Jones}{1991}]{Coles91}
Coles P.,  Jones B.,  1991, \mn@doi [MNRAS] {10.1093/mnras/248.1.1}, 248, 1

\bibitem[\protect\citeauthoryear{Crocce, Pueblas  \& Scoccimarro}{Crocce
  et~al.}{2006}]{Crocce2006}
Crocce M.,  Pueblas S.,   Scoccimarro R.,  2006, \mn@doi [MNRAS]
  {10.1111/j.1365-2966.2006.11040.x}, 373, 369

\bibitem[\protect\citeauthoryear{Croft \& Gaztanaga}{Croft \&
  Gaztanaga}{1998}]{Croft:1997rv}
Croft R. A.~C.,  Gaztanaga E.,  1998, \mn@doi [Astrophys. J.] {10.1086/305301},
  495, 554

\bibitem[\protect\citeauthoryear{{Cunha}, {Huterer}  \& {Dor{\'e}}}{{Cunha}
  et~al.}{2010}]{cunha/etal:2010}
{Cunha} C.,  {Huterer} D.,   {Dor{\'e}} O.,  2010, \mn@doi [\prd]
  {10.1103/PhysRevD.82.023004}, \href
  {http://adsabs.harvard.edu/abs/2010PhRvD..82b3004C} {82, 023004}

\bibitem[\protect\citeauthoryear{{DESI Collaboration} et~al.,}{{DESI
  Collaboration} et~al.}{2016}]{DESICollaboration2016a}
{DESI Collaboration} et~al., 2016, preprint (\mn@eprint {arXiv} {1611.00036})

\bibitem[\protect\citeauthoryear{Dalal, Dor\'e, Huterer  \& Shirokov}{Dalal
  et~al.}{2008}]{Dalal:2007cu}
Dalal N.,  Dor\'e O.,  Huterer D.,   Shirokov A.,  2008, \mn@doi [Phys. Rev.]
  {10.1103/PhysRevD.77.123514}, D77, 123514

\bibitem[\protect\citeauthoryear{De \& Croft}{De \& Croft}{2007}]{desoma}
De S.,  Croft R. A.~C.,  2007, \mn@doi [Mon. Not. Roy. Astron. Soc.]
  {10.1111/j.1365-2966.2007.11873.x}, \href
  {http://adsabs.harvard.edu/abs/2007MNRAS.382.1591D} {382, 1591}

\bibitem[\protect\citeauthoryear{De \& Croft}{De \& Croft}{2010}]{De:2009uz}
De S.,  Croft R. A.~C.,  2010, \mn@doi [Mon. Not. Roy. Astron. Soc.]
  {10.1111/j.1365-2966.2009.15804.x}, 401, 1989

\bibitem[\protect\citeauthoryear{Gangui, Lucchin, Matarrese  \&
  Mollerach}{Gangui et~al.}{1993}]{Gangui}
Gangui A.,  Lucchin F.,  Matarrese S.,   Mollerach S.,  1993, \mn@doi [ApJ]
  {10.1086/174421}, 430, 447

\bibitem[\protect\citeauthoryear{Gay, Pichon  \& Pogosyan}{Gay
  et~al.}{2012}]{Gay2012}
Gay C.,  Pichon C.,   Pogosyan D.,  2012, \mn@doi [PRD]
  {10.1103/PhysRevD.85.023011}, 85, 0

\bibitem[\protect\citeauthoryear{{Grinstein} \& {Wise}}{{Grinstein} \&
  {Wise}}{1986}]{grinstein/wise:1986}
{Grinstein} B.,  {Wise} M.~B.,  1986, \mn@doi [\apj] {10.1086/164660}, \href
  {http://adsabs.harvard.edu/abs/1986ApJ...310...19G} {310, 19}

\bibitem[\protect\citeauthoryear{Hamaus, Seljak, Desjacques, Smith  \&
  Baldauf}{Hamaus et~al.}{2010}]{Hamaus:2010im}
Hamaus N.,  Seljak U.,  Desjacques V.,  Smith R.~E.,   Baldauf T.,  2010,
  \mn@doi [Phys. Rev.] {10.1103/PhysRevD.82.043515}, D82, 043515

\bibitem[\protect\citeauthoryear{Kaiser \& N.}{Kaiser \&
  N.}{1984}]{Kaiser1984a}
Kaiser N.,  N. 1984, \mn@doi [ApJ] {10.1086/184341}, 284, L9

\bibitem[\protect\citeauthoryear{{Kamionkowski}, {Verde}  \&
  {Jimenez}}{{Kamionkowski} et~al.}{2009}]{kamionkowski:voids}
{Kamionkowski} M.,  {Verde} L.,   {Jimenez} R.,  2009, \mn@doi [\jcap]
  {10.1088/1475-7516/2009/01/010}, \href
  {http://adsabs.harvard.edu/abs/2009JCAP...01..010K} {1, 010}

\bibitem[\protect\citeauthoryear{Kofman, Bertschinger, Gelb, Nusser  \&
  Dekel}{Kofman et~al.}{1994}]{Kofman}
Kofman L.,  Bertschinger E.,  Gelb J.~M.,  Nusser A.,   Dekel A.,  1994,
  \mn@doi [ApJ] {10.1086/173541}, 420, 44

\bibitem[\protect\citeauthoryear{Komatsu \& Spergel}{Komatsu \&
  Spergel}{2001}]{KomatsuSpergel}
Komatsu E.,  Spergel D.~N.,  2001, \mn@doi [PRD] {10.1103/PhysRevD.63.063002},
  63, 13

\bibitem[\protect\citeauthoryear{{Lam}, {Desjacques}  \& {Sheth}}{{Lam}
  et~al.}{2010}]{lam/etal:2010}
{Lam} T.~Y.,  {Desjacques} V.,   {Sheth} R.~K.,  2010, \mn@doi [\mnras]
  {10.1111/j.1365-2966.2009.15903.x}, \href
  {http://adsabs.harvard.edu/abs/2010MNRAS.402.2397L} {402, 2397}

\bibitem[\protect\citeauthoryear{Laureijs et~al.,}{Laureijs
  et~al.}{2011}]{EuclidRB}
Laureijs R.,  et~al., 2011, preprint (\mn@eprint {arXiv} {1110.3193})

\bibitem[\protect\citeauthoryear{Lewis, Challinor  \& Lasenby}{Lewis
  et~al.}{2000}]{Lewis:1999bs}
Lewis A.,  Challinor A.,   Lasenby A.,  2000, \mn@doi [Astrophys. J.]
  {10.1086/309179}, 538, 473

\bibitem[\protect\citeauthoryear{{Ludlow} \& {Porciani}}{{Ludlow} \&
  {Porciani}}{2011}]{2011MNRAS.413.1961L}
{Ludlow} A.~D.,  {Porciani} C.,  2011, \mn@doi [\mnras]
  {10.1111/j.1365-2966.2011.18282.x}, \href
  {http://adsabs.harvard.edu/abs/2011MNRAS.413.1961L} {413, 1961}

\bibitem[\protect\citeauthoryear{{Marian}, {Hilbert}, {Smith}, {Schneider}  \&
  {Desjacques}}{{Marian} et~al.}{2011}]{marian/etal:2011}
{Marian} L.,  {Hilbert} S.,  {Smith} R.~E.,  {Schneider} P.,   {Desjacques} V.,
   2011, \mn@doi [\apjl] {10.1088/2041-8205/728/1/L13}, \href
  {http://adsabs.harvard.edu/abs/2011ApJ...728L..13M} {728, L13}

\bibitem[\protect\citeauthoryear{Matarrese \& Verde}{Matarrese \&
  Verde}{2008}]{Matarrese:2008nc}
Matarrese S.,  Verde L.,  2008, \mn@doi [Astrophys. J.] {10.1086/587840}, 677,
  L77

\bibitem[\protect\citeauthoryear{{Matsubara}}{{Matsubara}}{1994}]{matsubara:1994}
{Matsubara} T.,  1994, \mn@doi [\apjl] {10.1086/187570}, \href
  {http://adsabs.harvard.edu/abs/1994ApJ...434L..43M} {434, L43}

\bibitem[\protect\citeauthoryear{{Pillepich}, {Porciani}  \&
  {Reiprich}}{{Pillepich} et~al.}{2012}]{pillepich/etal:2012}
{Pillepich} A.,  {Porciani} C.,   {Reiprich} T.~H.,  2012, \mn@doi [\mnras]
  {10.1111/j.1365-2966.2012.20443.x}, \href
  {http://adsabs.harvard.edu/abs/2012MNRAS.422...44P} {422, 44}

\bibitem[\protect\citeauthoryear{{Planck Collaboration} et~al.,}{{Planck
  Collaboration} et~al.}{2016}]{PlanckPNG}
{Planck Collaboration} et~al., 2016, \mn@doi [\aap]
  {10.1051/0004-6361/201525836}, \href
  {http://adsabs.harvard.edu/abs/2016A%26A...594A..17P} {594, A17}

\bibitem[\protect\citeauthoryear{Press \& Schechter}{Press \&
  Schechter}{1974}]{PS}
Press W.~H.,  Schechter P.,  1974, \mn@doi [ApJ] {10.1086/152650}, 187, 425

\bibitem[\protect\citeauthoryear{Salopek \& Bond}{Salopek \&
  Bond}{1990}]{Salopek1990}
Salopek D.~S.,  Bond J.~R.,  1990, \mn@doi [PRD] {10.1103/PhysRevD.42.3936},
  42, 3936

\bibitem[\protect\citeauthoryear{Scoccimarro, Hui, Manera  \& Chan}{Scoccimarro
  et~al.}{2012}]{Scoccimarro12}
Scoccimarro R.,  Hui L.,  Manera M.,   Chan K.~C.,  2012, \mn@doi [PRD]
  {10.1103/PhysRevD.85.083002}, 85

\bibitem[\protect\citeauthoryear{Slosar, Hirata, Seljak, Ho  \&
  Padmanabhan}{Slosar et~al.}{2008}]{Slosar:2008hx}
Slosar A.,  Hirata C.,  Seljak U.,  Ho S.,   Padmanabhan N.,  2008, \mn@doi
  [JCAP] {10.1088/1475-7516/2008/08/031}, 0808, 031

\bibitem[\protect\citeauthoryear{Springel}{Springel}{2005}]{Gadget2}
Springel V.,  2005, \mn@doi [MNRAS] {10.1111/j.1365-2966.2005.09655.x}, 364,
  1105

\bibitem[\protect\citeauthoryear{Suginohara}{Suginohara}{1991}]{Suginohara1991}
Suginohara T.,  1991, ApJ, 371, 470

\bibitem[\protect\citeauthoryear{{Uhlemann}, {Pajer}, {Pichon}, {Nishimichi},
  {Codis}  \& {Bernardeau}}{{Uhlemann} et~al.}{2018}]{uhlemann/etal:2018}
{Uhlemann} C.,  {Pajer} E.,  {Pichon} C.,  {Nishimichi} T.,  {Codis} S.,
  {Bernardeau} F.,  2018, \mn@doi [\mnras] {10.1093/mnras/stx2623}, \href
  {http://adsabs.harvard.edu/abs/2018MNRAS.474.2853U} {474, 2853}

\bibitem[\protect\citeauthoryear{White \& Rees}{White \&
  Rees}{1978}]{White1978}
White S. D.~M.,  Rees M.~J.,  1978, \mn@doi [MNRAS] {10.1093/mnras/183.3.341},
  183, 341

\makeatother
\end{thebibliography}

\label{lastpage}
  
\end{document}